\documentclass[journal]{IEEEtran}

\ifCLASSINFOpdf
\else
   \usepackage[dvips]{graphicx}
\fi
\usepackage{url}

\usepackage{graphicx}
\usepackage{tabularx}
\usepackage{comment}
\usepackage{subfig}
\usepackage{svg}
\usepackage[justification=centering]{caption}
\usepackage{amsmath}
\usepackage{cite}
\usepackage{amsmath,amssymb,amsfonts}
\usepackage{algorithmic}
\usepackage{graphicx}
\usepackage{booktabs} 
\usepackage{multirow}
\usepackage{svg}
\usepackage{multicol}
\usepackage{comment}
\usepackage{caption} 
\usepackage[utf8]{inputenc} 
\usepackage{textcomp}
\usepackage{xcolor}
\begin{document}

\title{A Convolutional-Attentional Neural Framework for Structure-Aware Performance-Score Synchronization}

\author{Ruchit Agrawal,~\IEEEmembership{Student Member,~IEEE} $^{\dagger}$, Daniel Wolff $^{\star}$, and Simon Dixon $^{\dagger}$  \\
$^{\dagger}$ Centre for Digital Music, Queen Mary University of London, UK \\
      $^{\star}$ Institute for Research and Coordination in Acoustics/Music, Paris, France
\thanks{This manuscript was accepted to IEEE SPL on 18-11-2021. This research is supported by the European Union’s Horizon 2020 research and innovation programme under the Marie
Skłodowska-Curie grant agreement No. 765068.}
\thanks{Ruchit Agrawal (e-mail: r.r.agrawal@qmul.ac.uk) and Simon Dixon (e-mail: s.e.dixon@qmul.ac.uk) are with the Centre for Digital Music, Queen Mary University of London, Mile End Road, London E1 4NS, United Kingdom.}
\thanks{Daniel Wolff (e-mail: daniel.wolff@ircam.fr) is with the Institute for Research and Coordination in Acoustics/Music, 75004 Paris, France.}}
\markboth{IEEE Signal Processing Letters, November 2021}
{Shell \MakeLowercase{\textit{et al.}}: Bare Demo of IEEEtran.cls for IEEE Journals}

\maketitle
\begin{abstract}
Performance-score synchronization is an integral task in signal processing, which entails generating an accurate mapping between an audio recording of a performance and the corresponding musical score. Traditional synchronization methods compute alignment using knowledge-driven and stochastic approaches, and are typically unable to generalize well to different domains and modalities. We present a novel data-driven method for structure-aware performance-score  synchronization. We propose a convolutional-attentional architecture trained with a custom loss based on time-series divergence. We conduct experiments for the audio-to-MIDI and audio-to-image alignment tasks pertained to different score modalities. We validate the effectiveness of our method via ablation studies and comparisons with state-of-the-art alignment approaches. We demonstrate that our approach outperforms previous synchronization methods for a variety of test settings across score modalities and acoustic conditions. Our method is also robust to structural differences between the performance and score sequences, which is a common limitation of standard alignment approaches.
\end{abstract}
\vspace{-0.1cm}

\IEEEpeerreviewmaketitle

\vspace{-0.1cm}
\section{Introduction}

\IEEEPARstart{T}{he} alignment of time-series data corresponding to different sources of information is an important task in signal processing, with applications in a variety of scenarios such as subtitle synchronization, performance analysis and speech recognition. Performance-score synchronization is one such alignment task that is aimed at computing the optimal mapping between a performance and the score for a given piece of music. Depending upon the task, the alignment computation is either carried out online, known as \emph{score following}, or offline, known as \emph{performance-score synchronization} or \emph{audio-to-score alignment}. In this paper, we focus on the offline alignment task, i.e. \emph{performance-score synchronization}.  Traditional alignment methods are generally based on Dynamic Time Warping (DTW) \cite{dixon2005line} or Hidden Markov Models (HMMs) \cite{muller2015fundamentals}. 
A particular limitation of DTW-based methods is the inability to capture structural differences between the two input sequences, since the alignments computed using DTW are constrained to progress monotonically through both the sequences. Similarly, the Markov assumption in HMMs limits contextual modelling, which has proven to be useful in various ML tasks \cite{amirhossein2018multi, agrawal2018contextual, agrawal2018no}.
Data-driven approaches have shown promise for various signal processing tasks, including alignment \cite{dorfer2018learning2}, \cite{agrawal2021structure}.  Recent methods have proposed the use of neural networks for the similarity computation \cite{dorfer2017learning, agrawal2021learning}, coupled with DTW for the alignment computation. 
This paper furthers the development of data-driven alignment approaches and proposes a neural architecture for \textit{learnt} alignment computation, thereby eschewing the limitations of DTW-based alignment. 

\par We present a novel neural method for performance-score synchronization, which is also robust to structural differences between the performance and the score.
We propose a convolutional-attentional encoder-decoder architecture, with the encoder based on a convolutional stem and the decoder based on a stand-alone self-attention block \cite{ramachandran2019stand}. 
Our motivation behind employing the stand-alone self-attention block, as opposed to the more commonly used approach of an attention computation on top of the convolution operation, is that the stand-alone self-attention layers have proven to be effective at capturing global relations in vision tasks when employed in later stages of a convolutional network \cite{ramachandran2019stand}. In addition to proposing a novel architecture, we employ a customized divergence loss function based on the differentiable soft-DTW computation \cite{cuturi2017soft} to train our models. We conduct experiments for two synchronization tasks involving different score modalities, namely audio-to-MIDI alignment, i.e. aligning audio to symbolic music representations, and audio-to-image alignment, i.e. aligning audio to scanned images of sheet music. Our results demonstrate that the proposed method generates robust alignments in both settings.
The primary contributions of this paper are summarized below:
\begin{itemize}
    \item We present a novel neural architecture for learning structure-aware performance-score synchronization.
    \item We demonstrate that combining stand-alone self-attention layers with a convolutional stem outperforms multiple alignment methods across different test settings.
    \item We demonstrate that our method effectively handles performances containing structural deviations from the score. 
    \item We demonstrate that the custom soft-DTW based divergence is an effective loss function for training performance-score synchronization models.
\end{itemize}

\begin{figure*}[th]
\vspace{-0.1cm}
  \centering
\includegraphics[width=6.5in]{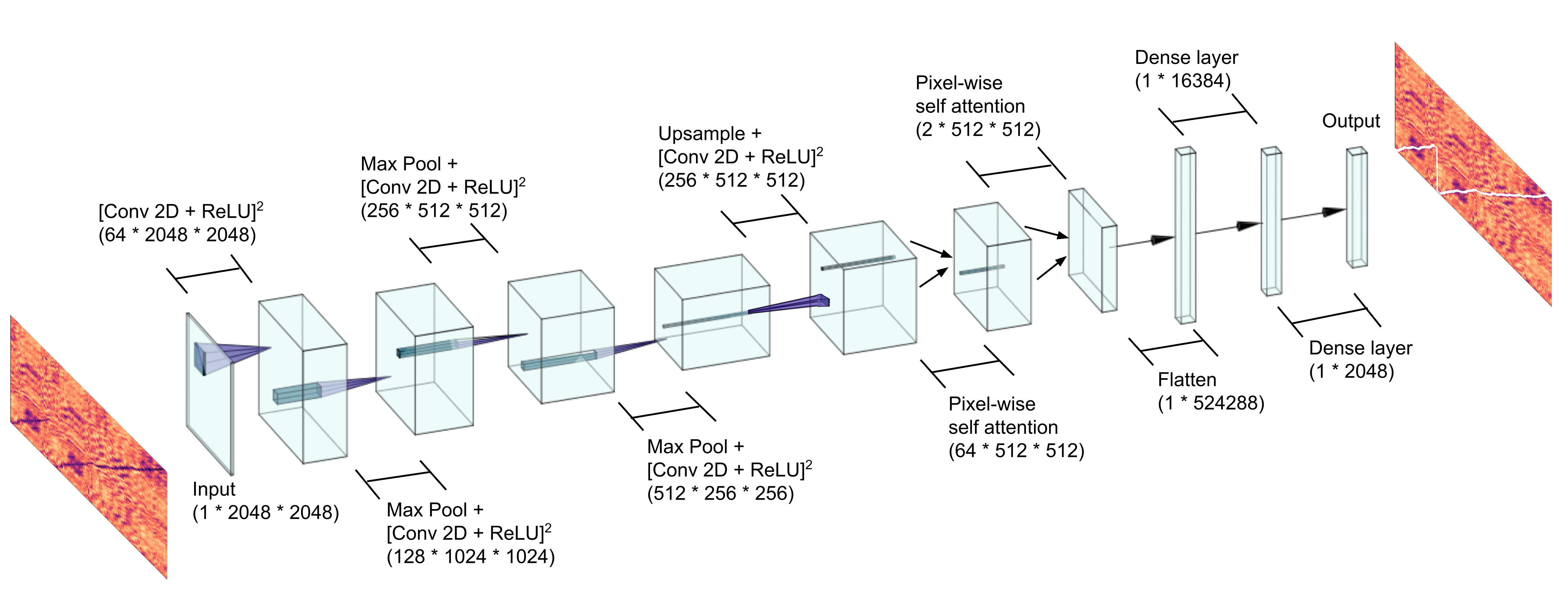}
  \caption{Schematic diagram illustrating our model architecture. \\ The output alignment path is plotted against the distance matrix for a simple example to aid visualization.} 
  \vspace{-0.4cm}
  \label{fig:pipeline}
\end{figure*}

\section{Related Work}\label{related}
\vspace{-0.1cm}
Traditional performance-score synchronization approaches relied mainly on DTW \cite{muller2004towards,  dixon2005line, muller2006efficient, arzt2008automatic} or HMMs \cite{cont2005training, maezawa2011polyphonic, cuvillier2014coherent, gong2015real}. Methods were subsequently proposed that modified DTW for optimizing alignment performance across various settings \cite{zhou2009canonical, arzt2010towards, carabias2015audio, wang2016robust}.
With the advent of data-driven approaches, recent methods have demonstrated the efficacy of learnt representations coupled with DTW-based alignment computation for performance synchronization \cite{dorfer2017learning, agrawal2020hybrid, tanprasert2019midi, agrawal2021learning}. A step further are fully learnt methods, which have been explored for generic multiple sequence alignment. These include Deep Canonical Time Warping \cite{trigeorgis2017deep}, the first deep temporal alignment method for simultaneous feature selection and alignment; NeuMATCH \cite{Dogan18neumatch}, an alignment method based on LSTM blocks, and more recently, Neural Time Warping \cite{kawano2020neural} that models multiple sequence alignment as a continuous optimization problem. 
While neural methods have recently been explored for score following \cite{dorfer2018learning, henkel2020learning}, their application to performance-score synchronization remains relatively unexplored.

\par An approach related to our method is the audio-conditioned U-net for position estimation in sheet images \cite{henkel2019audio}, later adapted for score following by addressing the temporal aspects of the task \cite{henkel2020learning}.
While they employ a U-net architecture and focus on score following, we present a hybrid convolutional-attentional architecture and focus on performance-score synchronization. We also employ a custom loss to train our models. Additionally, our method is able to handle structural deviations from the score, which is a limitation of their model. 
Approaches to specifically tackle structural differences in audio-to-score alignment include \begin{math}\mathit{JumpDTW}\end{math} \cite{Fremerey2010handling}, Needleman-Wunsch Time Warping  \cite{grachten2013automatic} (\begin{math}\mathit{NWTW}\end{math}) and progressively dilated convolutional neural networks \cite{agrawal2021structure}. Our method differs from these in that we do not explicitly model the structural deviations, and our architecture learns to model them inherently while predicting the alignment. Additionally, the alignment computation is performed by our network itself, rather than relying on a dynamic programming framework.
\vspace{-0.2cm}
\section{Proposed Method}\label{sec:method}
We model the performance-score synchronization task as a sequence prediction task, given the two input sequences corresponding to the performance and score respectively. However, rather than relying on recurrent neural networks or Transformers \cite{vaswani2017attention} and predicting the output sequence one token at a time, we propose a convolutional-attentional architecture that predicts the entire alignment path in a one-shot fashion. This allows the model to capture long-term dependencies and also handle structural differences between the performance and score sequences.
The architecture of our model is depicted in Figure \ref{fig:pipeline}. Our network operates on the cross-similarity matrix between the performance and score feature sequences and predicts the ($x$, $y$) co-ordinates corresponding to the frame indices that make up the optimal alignment path. Since the $X$-axis of the matrix corresponds to the performance, it progresses linearly in the alignment and the goal is essentially to predict the sequence of $y$ co-ordinates (i.e. frame indices in score axis) that determine the alignment path. Formally, let $X$= $(x_1, x_2,..., x_p)$ and $Y$ = $(y_1, y_2,..., y_q)$ be the feature sequences corresponding to the performance and score respectively. The network is trained to predict the sequence of frame indices $\hat{Y}_p$= $(\hat{y}_1, \hat{y}_2,..., \hat{y}_p)$, denoting the path taken by the performance $X$ through the score $Y$. 

\vspace{-0.2cm}
\par Our network has an encoder-decoder architecture, with the encoder comprising four convolutional and downsampling blocks, and the decoder comprising an upsampling block, a stand-alone self-attention (hereafter abbreviated as \emph{SASA}) block  and a fully connected block. It must be noted that the \emph{SASA} block employed by our decoder is different from the commonly explored combination of an attention computation applied on top of a convolution operation \cite{oktay2018attention}, or the self-attention layer from the sequence to sequence Transformer architecture\cite{vaswani2017attention}.
The \emph{SASA} block borrows ideas from both convolution and self-attention, and is able to replace spatial convolutions completely and effectively integrate global information, especially when deployed in the later stage of a convolutional neural network \cite{ramachandran2019stand}. 
As part of our upsampling strategy, we employ the max-unpooling operation \cite{zeiler2014visualizing} as opposed to the transposed convolution, which has been shown to result in artifacts \cite{odena2016deconvolution}. We store the indices of the highest activations during pooling and pass these recorded locations to the upsampling block, where the max-unpooling places each element in the unpooled map according to the mask, instead of assigning it to the upper-left pixel.

\par The upsampled output is passed on to the attentional block, comprising two \emph{SASA} layers. For each pixel $(i, j)$ in the upsampled output, we compute the self-attention relative to the memory block $M_{k}(i, j)$, which is a neighbourhood with spatial extent $k$ centered around $(i, j)$, as follows:
\begin{equation}
y_{ij} = \hspace{-0.2cm} \sum\limits_{a, b \in M_{k}(i, j)} \hspace{-0.4cm}\mathit{softmax_{ab}}  (q_{ij}^ \intercal k_{ab} + q_{ij}^ \intercal r_{a-i, b-j}) \hspace{0.1cm}  v_{ab}
\end{equation}
where $q_{ij} = W_q x_{ij}$ are the queries, $k_{ab} = W_k x_{ab}$ the keys and $v_{ab} = W_v x_{ab}$ the values computed as linear transformations from the activations at the $(i, j)^{th}$ pixel and its memory block. The displacements from the current position $(i, j)$ to the neighborhood pixel $(a, b)$ are encoded by row and column offsets, given by $r_{a-i}$ and $r_{b-j}$ respectively. We employ four attention heads and split the pixel features depthwise into four groups of the same size.  The attention is then computed on each group individually with different matrices $W$ and the results are concatenated to yield the pixel-wise attention values $y_{ij}$. This computation is repeated twice and the output is passed through a fully connected block with two dense layers to predict the alignment path. A graphic elaboration can be found in Figure 2 in the supplementary material.
 
\par We employ a time-series divergence loss function to train our models, as opposed to using a cross entropy loss. The primary motivation behind this loss is that it allows us to minimize the overall cost of aligning the performance and score feature sequences by comparing the \textit{paths} rather than the \textit{feature sequences} using a positive definite divergence.
Our loss function captures the divergence between the predicted and ground truth alignment sequences, based on the soft-DTW \cite{cuturi2017soft} distance. We employ soft-DTW since it offers a differentiable measure of the discrepancy between the two sequences. Given the predicted alignment sequence $\hat{Y}$= $(\hat{y}_1, \hat{y}_2,..., \hat{y}_p)$ and the ground truth alignment sequence $Y$ = $(y_1, y_2,..., y_r)$, we compute the soft-DTW distance $D_\lambda (p, r)$ as follows:
\begin{equation}\label{eq:softDTW}
    D_\lambda (p, r) = e(p, r) + min_{\lambda} \begin{cases}
    D_\lambda (p, r-1) \\ D_\lambda (p-1, r) \\  D_\lambda (p-1,  r-1) \\
    \end{cases}
\end{equation}
where $e(p, r)$ is the Euclidean distance between points $\hat{y}_p$ and $y_r$, and $min_{\lambda}$ is the soft-min operator parametrized by a smoothing factor \begin{math}\lambda \end{math}, as follows: \\
\begin{equation}
    min_{\lambda}\{m_1, m_2, ..., m_n\} = \begin{cases}
min\{m_1, m_2, ..., m_n\} 
\hspace{0.5cm} \lambda=0 \\ -\lambda \log \sum_{i=1}^{i=n} e^{-m_i/\lambda} \\
\end{cases}
\end{equation}

We then normalize $D_\lambda (p, r)$ in order to make it a 
positive definite divergence \cite{blondel2021differentiable}, as follows:  
\begin{equation} \label{eq:divergence}
SD_\lambda (\hat{Y}, Y) = D_\lambda (\hat{Y}, Y) - 1/2(D_\lambda (\hat{Y}, \hat{Y}) + D_\lambda (Y, Y))
\end{equation}
This ensures that $SD_\lambda(\hat{Y}, Y)>0$ for $\hat{Y}\neq Y$  and $SD_\lambda(\hat{Y}, Y) = 0$ for $\hat{Y} = Y$, yielding a completely learnable framework since $SD_\lambda(\hat{Y}, Y)$ is non-negative and differentiable at all points. Note that we only employ the distance metric $D_\lambda (p, r)$ from the soft-DTW computation in Equation \ref{eq:softDTW}, and not the alignment path itself. The alignment computation is carried out by our neural framework, by minimizing the custom divergence loss $SD_\lambda(\hat{Y}, Y)$. 
\vspace{-0.2cm}
\section{Experiments and Results}\label{experiments}
\vspace{-0.1cm}

We conduct experiments for two alignment tasks pertinent to different score modalities, namely audio-to-MIDI alignment and audio-to-image alignment.
For each performance-score pair, the cross-similarity matrix is computed using the Euclidean distance between the chromagrams for audio-to-MIDI alignment, and the Euclidean distance between learnt cross-modal embeddings \cite{dorfer2017learning} for audio-to-image alignment. We employ librosa \cite{mcfee2015librosa} for computing the chromagrams and the cross-similarity matrices.  We employ a sampling rate of 22050 Hz, a frame length of 2048 samples and a hop length of 512 samples for the chromagram computation.
On the encoder side, the output of each 2D convolution is batch normalized and passed through a Rectified Linear Unit (ReLU) non-linearity, before being passed on to max-pooling.
We employ a dropout of 0.4 for the fully connected layers to avoid overfitting. The output of the final layer is a vector of length 2048, encoding the y-indices making up the alignment path. During training and testing, each performance feature sequence is scaled to length 2048, and the score feature sequence is padded accordingly. These are then rescaled back to the original dimensions for comparing the predicted alignment with the ground truth.
It must be noted that the output vector is sufficient to capture the length of all pieces in the data, since the audio-to-MIDI task has beat-level annotations (less than 2048 per piece), and the audio-to-image task has notehead-level annotations (also less than 2048 per piece). 
Since the entire vector is predicted in a one-shot manner, we do not encounter instability issues and hence do not explore smoothing for refining the predicted alignment path. 
\par We employ two publicly available datasets for our experiments on the two alignment tasks. For the audio-to-MIDI alignment task, we employ the  Mazurka-BL dataset \cite{kosta2018mazurkabl}. This dataset comprises 2000 recordings with annotated alignments at the beat level. The recordings correspond to performances of Chopin’s Mazurkas dating from 1902 to the early 2000s, and
 span various acoustic settings. We randomly divide this set into sets of 1500, 250 and 250 recordings respectively, forming the training, validation and testing sets. 
We compare the results obtained by our models with \begin{math} \textit{MATCH}\end{math} \cite{dixon2005line}, \begin{math}\textit{JumpDTW}\end{math} \cite{Fremerey2010handling}, \begin{math}\textit{SiameseDTW}\end{math} \cite{agrawal2021learning},  \begin{math}\textit{DilatedCNN}\end{math} \cite{agrawal2021structure}, and the \begin{math}\textit{Deep Canonical Time Warping}\end{math} \emph{(DeepCTW)} method \cite{trigeorgis2017deep}. We also conduct ablative studies to analyze the specific improvements obtained by employing \emph{SASA} and the custom loss function in our architecture. To this end, we replace the \emph{SASA} layers in our model with convolutional layers keeping the input and output dimensionalities constant. The resulting Conv-Deconv models are abbreviated as \emph{CD}$_{x}$, with $x$ denoting the loss function employed, i.e. \emph{CE} for the cross-entropy loss and \emph{custom} for the custom loss. Our convolutional-attentional models are similarly abbreviated as \emph{CA}$_{x}$. 
We compute the percentage of beats aligned correctly within error margins of 50, 100 and 200 ms respectively for each piece, and report the alignment accuracy \cite{cont2007evaluation} obtained by each model averaged over the entire test set in Table I. We also conduct significance testing using the Diebold-Mariano test \cite{harvey1997testing} and perform pairwise comparisons of all model predictions with the \emph{CA}$_{\textit{custom}}$ predictions for each error margin (Table I).

\par For the audio-to-image alignment task, we employ the Multimodal Sheet Music Dataset \cite{dorfer2018learning}, a standard dataset  for sheet image alignment analysis. It comprises polyphonic piano music for 495 classical pieces, with notehead-level annotations linking the audio files to the sheet images.  
We divide this set randomly into sets of 400, 50 and 45 recordings respectively, forming the training, validation and testing sets. We compare the results obtained by our model with contemporary audio-to-image alignment methods \begin{math} \textit{Dorfer et al.}\end{math} \cite{dorfer2017learning},  \cite{dorfer2018learning2}, 
\begin{math} \textit{Henkel et al.}\end{math} \cite{henkel2020learning}, and the  
\begin{math}\textit{DilatedCNN}\end{math} model \cite{agrawal2021structure}. 
For comparison with Henkel et al. \cite{henkel2020learning}, we interpolate their predicted sheet image co-ordinates to the time domain from the ground truth alignment between the note onsets and the corresponding notehead co-ordinates in the sheet images. We then compute the alignment accuracy obtained by each model by calculating the percentage of onsets aligned correctly within the error margins of 500 ms, 1 s and 2 s respectively, and report the results averaged over the test set along with the significance tests in Table II. Note that we employ the same feature representation \cite{dorfer2017learning} for all audio-to-image methods. 

\par In addition to the overall accuracy on the test sets, we also report alignment results for structurally different performance-score pairs for both datasets. In order to specifically test the model performance on structure-aware alignment for both the tasks, we generate 20\% additional samples that contain structural differences between the score and the performance. These are generated using a randomized split-join operation using the audio from the respective datasets. We append 50\% of these samples to our training sets and employ the other 50\% as our testing sets. The ground truth alignments are extrapolated from the original alignments using the split-join locations. The two best performing models for each evaluation setup are highlighted in Tables I and II.
\begin{table}[t]
\vspace{-0.2cm}
   \centering
\begin{tabular}{cccccc} \toprule
\hline 
\multirow{2}{*}{\textbf{Model}} & 
\multicolumn{3}{c}{\textit{Overall}} 
& \textit{Structure}
\tabularnewline
  & \textbf{$<$50 ms}& \textbf{$<$100 ms} & \textbf{$<$200 ms} & \textbf{$<$100 ms} 
  \\
\midrule 
 \begin{math}\textit{MATCH}\end{math}\cite{dixon2005line} &  74.6* & 79.5* & 85.2* & 67.4* \\
\midrule

 \begin{math}\textit{JumpDTW}\end{math}\cite{Fremerey2010handling} &  75.2* & 80.4* & 86.7* & 76.2* \\
\midrule

 \emph{SiameseDTW} \cite{agrawal2021learning}  & \underline{77.9} & 83.3* & 89.5* & 72.8*  \\
\midrule 
  \begin{math}\textit{DeepCTW}\end{math}\cite{trigeorgis2017deep}  & 76.1* & 81.6* & 88.9* & 75.6* \\
\midrule 
 \begin{math}\textit{DilatedCNN}\end{math}\cite{agrawal2021structure} & 77.5* & 82.4* & 90.4* & \textbf{80.3}  \\
\midrule 
  \emph{CD$_{\textit{CE}}$} & 72.8* & 80.1* & 85.3* & 71.9*  \\
  \midrule 
  \emph{CD$_{\textit{custom}}$} & 74.1* & 81.7* & 87.5* & 74.2*  \\
  \midrule 
  \emph{CA$_{\textit{CE}}$} & 76.4* & \underline{84.1}* & \underline{90.9}* & 76.8*  \\
  \midrule
  \emph{CA$_{\textit{custom}}$} & \textbf{78.7} & \textbf{85.2} & \textbf{92.6} & \underline{79.5}  \\
\midrule 
\bottomrule
\end{tabular}
\caption{Audio-to-MIDI alignment accuracy in \% on the \emph{Mazurka-BL} dataset. Best in bold, second best underlined.\\$*$: significant differences from \emph{CA$_{\textit{custom}}$}, $p < 0.05$}
\vspace{-0.5cm}
\label{results_score}
\end{table}

\vspace{-0.2cm}
\section{Discussion and Conclusion}\label{discussion}
The results obtained by our method for audio-to-MIDI alignment are given in Table \ref{results_score}.
Overall accuracy on the Mazurka-BL dataset suggests that our best model outperforms the DTW-based frameworks \emph{MATCH}, \emph{JumpDTW} and \emph{SiameseDTW} by up to 7\% (Table I, rows 1-3) as well as the neural frameworks \begin{math}\mathit{DeepCTW}\end{math} and \begin{math}\mathit{Dilated} \mathit{CNN}\end{math}) by up to 4\% (Table I, rows 4-5) for all error margins. 
 The comparison with contemporary approaches reveals that our method yields higher improvement over the state-of-the art for coarse alignment (Error margins $>$ 50ms) than for fine-grained alignment 
 (Error margin $<$ 50ms). Moreover, the ablative studies suggest that the convolutional-attentional architecture (\emph{CA}) outperforms the conv-deconv architecture (\emph{CD}) by 3-5\%, with higher improvement (4-6\%) observed on structurally different pieces (Table I, column 4). Additionally, the custom loss yields an improvement of 1-3\% over the cross-entropy loss, for both the \emph{CD} and \emph{CA} architectures (Table I, rows 6-9), with the \emph{CA}$_{\textit{custom}}$ model yielding the best overall performance.
The experimentation for audio-to-image alignment similarly  reveals that \emph{CA}$_{\textit{custom}}$ outperforms Dorfer et al. \cite{dorfer2017learning, dorfer2018learning2} and  \begin{math}\textit{DilatedCNN}\end{math}  \cite{agrawal2021structure} in overall alignment accuracy by 2-10\% and Henkel et al. \cite{henkel2020learning} by 1-4\% for all error margins (Table II, columns 1-3). A further advantage of our method over Henkel et al. \cite{henkel2020learning} is the ability to work with pieces containing several pages of sheet music, as opposed to only one. Our model is also able to handle structural deviations from the score, which is a limitation of the majority of alignment approaches, including Henkel et al. \cite{henkel2020learning} (Table II, column 4). 

\begin{table}[t]
\vspace{-0.4cm}
   \centering
\begin{tabular}{ccccc} \toprule
\hline 
\multirow{2}{*}{\textbf{Model}} & 
\multicolumn{3}{c}{\textit{Overall}} &
\textit{Structure}
\tabularnewline
  & \textbf{$<$0.5 s}& \textbf{$<$1 s} & \textbf{$<$2 s} & \textbf{$<$1 s} 
  \\
\midrule 
 \begin{math}\textit{Dorfer et al. 2017}\end{math}\cite{dorfer2017learning} & 73.5* & 81.2* & 84.7* & 67.8*  \\
\midrule
 \begin{math}\textit{Dorfer et al. 2018}\end{math} \cite{dorfer2018learning2} & 76.4* & 84.5* & 89.3* & 70.3*   \\
\midrule 
 \begin{math}\textit{DilatedCNN}\end{math}\cite{agrawal2021structure}  & 82.8* & 87.6* & \underline{90.8}* & \textbf{78.5}  \\
\midrule 
   \begin{math}\textit{Henkel et al. 2020}\end{math} \cite{henkel2020learning}  & \underline{84.6} & \underline{88.4}* & 90.1* & 72.1* \\
\midrule 
  \emph{CA$_{\textit{custom}}$}  & \textbf{85.2} & \textbf{91.5} & \textbf{92.9} & \underline{77.4}  \\
\midrule 
\bottomrule
\end{tabular}
\caption{Audio-to-Image alignment accuracy in \% on the \emph{MSMD} dataset. Best in bold, second best underlined.\\$*$: significant differences from \emph{CA$_{\textit{custom}}$}, $p < 0.05$}  \label{results_image}
\vspace{-0.5cm}
\end{table}
\par Experiments specifically on structure-aware alignment demonstrate that our method outperforms all approaches except \begin{math}\textit{DilatedCNN}\end{math} \cite{agrawal2021structure} for both the tasks by 3-12\% (Tables 1, 2, column 4).
Our model \emph{CA}$_{\textit{custom}}$ demonstrates comparable results to \begin{math}\textit{DilatedCNN}\end{math} \cite{agrawal2021structure}, without explicitly modeling structure, and while being trained on limited structure-aware data. The ablative analysis demonstrates that the \emph{CA}$_{x}$ models yield better structure-aware alignment than the \emph{CD}$_{x}$ models (Table I, rows 6-9, column 4), confirming that the stand-alone self-attention layers in the decoder facilitate long-term contextual incorporation. Manual inspection of the alignment plots corroborated that \emph{CA}$_{\textit{custom}}$ was able to capture structural deviations such as jumps and repeats. The reader can find such examples in Figure 3 in the supplementary material.  
\par To conclude, we demonstrate that the proposed convolutional-attentional architecture trained with a custom time-series divergence loss is a promising framework for performance-score synchronization. Our approach is compatible with both multi-modal and uni-modal data, since the similarity and alignment computations are done separately. 
Our method is also robust to structural differences between the performance and score sequences without explicit structure modeling.
In the future, we would like to explore multi-modal methods that work directly with raw data, and dynamic neural methods that can adjust to the alignment granularity needed for the task at hand. 
\bibliographystyle{IEEEbib}

\bibliography{refs}

\end{document}